\documentclass[aip,preprint]{revtex4-2}
\usepackage[cp1251]{inputenc}
\usepackage[T2A]{fontenc}
\usepackage[english]{babel}
\usepackage{amssymb,latexsym,amsmath,amscd}
\usepackage{graphicx,color,framed}
\usepackage{tikz}
\usepackage{xcolor}
\usepackage{siunitx}
\usepackage{empheq}
\usepackage{xr}
\usepackage{microtype}
\usepackage{amsfonts}
\usepackage[normalem]{ulem}
\graphicspath{{figures/}}

\makeatletter
\newcommand*{\addFileDependency}[1]{
  \typeout{(#1)}
  \@addtofilelist{#1}
  \IfFileExists{#1}{}{\typeout{No file #1.}}
}
\makeatother

\begin{document}
\newcommand{\solid}{\raisebox{2pt}{\tikz\draw[solid](0,0) -- (5mm,0);}}

\title{Theory of self-coacervation in semi-dilute and concentrated zwitterionic polymer solutions}
\author{\firstname{Yury A.} \surname{Budkov}}
\email[]{ybudkov@hse.ru}
\affiliation{School of Applied Mathematics, HSE University, Tallinskaya st. 34, 123458 Moscow, Russia}
\affiliation{G.A. Krestov Institute of Solution Chemistry of the Russian Academy of Sciences, 153045, Akademicheskaya st. 1, Ivanovo, Russia}
\author{\firstname{Petr E.} \surname{Brandyshev}}
\affiliation{School of Applied Mathematics, HSE University, Tallinskaya st. 34, 123458 Moscow, Russia}
\author{\firstname{Nikolai N.} \surname{Kalikin}}
\affiliation{School of Applied Mathematics, HSE University, Tallinskaya st. 34, 123458 Moscow, Russia}
\affiliation{G.A. Krestov Institute of Solution Chemistry of the Russian Academy of Sciences, 153045, Akademicheskaya st. 1, Ivanovo, Russia}

\begin{abstract}
Based on the random phase approximation, we develop a molecular theory of self-coacervation in zwitterionic polymer solutions. We show that the interplay between the volume interactions of the monomeric units and electrostatic correlations of charged groups on a polymer backbone can result in liquid-liquid phase separation (self-coacervation). We analyse the behavior of the coacervate phase polymer concentration depending on the electrostatic interaction strength -- the ratio of the Bjerrum length to the bond length of the chain. We establish that in a wide range of polymer concentration values -- from a semi-dilute to a rather concentrated solution -- the chain connectivity and excluded volume interaction of the monomeric units have an extremely weak effect on the contribution of the electrostatic interactions of the dipolar monomeric units to the total free energy. We show that for rather weak electrostatic interactions, the electrostatic correlations manifest themselves as Keesom interactions of point-like freely rotating dipoles (Keesom regime), while in the region of strong electrostatic interactions the electrostatic free energy is described by the Debye-H{\"u}ckel limiting law (Debye regime). We show that for real zwitterionic coacervates the Keesom regime is realized only for sufficiently small polymer concentrations of the coacervate phase, while the Debye regime is approximately realized for rather dense coacervates. Using the mean-field variant of the density functional theory, we calculate the surface tension (surface free energy) of the $"$coacervate-solvent$"$ interface as a function of the bulk polymer concentration.  {Obtained results can be used to estimate the parameters of the polymer chains needed for practical applications such as drug encapsulation and delivery, as well as the design of adhesive materials.}
\end{abstract}

\maketitle
\section{Introduction}
Coacervation plays an extremely important role in macromolecular assembly in soft liquid-phase media with different functions. Coacervation is the result of spontaneous separation of a macromolecular solution into two immiscible liquid phases of low and high concentration -- liquid-liquid phase separation. The latter is a common physical phenomenon, occurring in aqueous solutions of polyelectrolytes, surfactants, and biomacromolecules -- systems that nowadays are extensively explored, both theoretically and experimentally, and have a number of applications in food, pharmaceutical, cosmetic, and chemical industries~\cite{xiao2022small}. On the other hand, in the 1930s, Alexander Oparin proposed the concept of coacervation in solutions of biomacromolecules underlying the origin of life~\cite{oparin1957origin}. And it was not until 2009 that biophysicists established the role of liquid-liquid phase separation in the formation of certain membraneless organelles~\cite{brangwynne2009germline}.  {Current applications of coacervates include wastewater treatment, protein purification, food processing, drug delivery, cosmetics formulation and synthesis of nanoparticles \cite{moulik2022overview}. In the food industry, coacervates have been used to create encapsulated flavors, fragrances, and nutrients that are released in a controlled manner during consumption~\cite{schmitt2011protein}. They can also be used as a natural preservative, preventing spoilage and increasing the shelf-life of products~\cite{devi2017encapsulation,schmitt2011protein}. Coacervates have also been applied in the textile industry, serving as a dyeing agent that enables the color to bind more strongly to the fabric. This leads to a more eco-friendly production process, as well as reduced chemical waste~\cite{timilsena2019complex,valley2019rapid}. Furthermore, the study of coacervates has provided insight into the self-assembly of biological molecules, such as proteins and nucleic acids. It has also led to the development of new technologies such as artificial cells, which have potential applications in drug delivery and synthetic biology~\cite{mashima2022dna,cook2023complex}.} Overall, the widespread relevance of coacervates has led to fundamentally new functional biomaterials and created a new wave of modern materials, extending the boundaries of polymer physics and chemistry~\cite{sing2020recent}.

These days, the theory of coacervation is being extensively developed for solutions of polyelectrolytes~\cite{borue1990statistical,kudlay2004complexation,ermoshkin2003modified,budkov2015new,shen2018polyelectrolyte,sing2020recent}, polyampholytes~\cite{rumyantsev2022unifying}, and intrinsically disordered proteins~\cite{lin2017random,mccarty2019complete}, in which phase separation is driven by strong electrostatic correlations of ionic groups. Nevertheless, there are a number of macromolecular systems, whose thermodynamic behavior is determined by the dipole-dipole interactions of highly polar or polarizable monomeric units~\cite{martin2016statistical}. The ones of utmost importance are macromolecules valuable for materials science applications such as zwitterionic polymers, whose monomeric units carry two oppositely charged ionic groups~\cite{peng2016zwitterionic,lei2018zwitterionic,li2020functional}.

One of the most interesting effects driven by dipole-dipole interactions is the coil-globule transition of a single dipolar polymer chain~\cite{gordievskaya2018interplay,cherstvy2010collapse,schiessel1998counterion,kumar2009theory}. The latter can occur in dilute solutions of the aforementioned zwitterionic polymers~\cite{kumar2009theory,gordievskaya2018interplay,budkov2021molecular}. However, in a dilute dipolar polymer solution sufficiently strong dipole-dipole interactions provoke a coil-globule transition of an individual polymer chain, while in semi-dilute and concentrated solutions such interactions should lead to a liquid-liquid phase separation (so-called self-coacervation). For the aforementioned industrial applications, it is important to study the phase behavior of such zwitterionic polymer solutions in the semi-dilute and concentrated regimes, where the excluded volume interactions are expected to interplay with the electrostatic correlations of the dipolar monomeric units. The latter, for example, could be used to control the coacervate density (or size of the coacervate droplets) by varying the electrostatic interaction strength in rather narrow ranges (by temperature or solvent composition). {We would like to note that recently, it has been experimentally observed that zwitterionic polymers such as sulfobetaines undergo an enthalpy-driven liquid-liquid phase separation in aqueous solutions, resulting in the formation of liquid coacervate droplets~\cite{capasso2022programmable,paganini2022high}.}

However, to the best of our knowledge, there are only a few theoretical studies devoted to the investigation of the coacervation effect in polymer solutions, induced by the dipole-dipole attraction of the macromolecule polar mononeric units. In paper~\cite{adhikari2018polyelectrolyte} the authors addressed complex coacervation -- a liquid-liquid phase separation of a solution of oppositely charged polyelectrolyte chains into a polyelectrolyte-rich complex coacervate phase and a dilute aqueous phase, based on the general premise of spontaneous formation of polycation-polyanion complexes. The complexes are treated as flexible chains consisting of dipolar and charged segments. Using a mean-field theory accounting for the entropy of all dissociated ions in the system, electrostatic interactions between the dipolar and charged segments of the complexes and separated polyelectrolytes, and polymer-solvent hydrophobicity, the authors predicted closed liquid-liquid phase diagrams with lower and upper critical points of such a system in terms of the polyelectrolyte composition, added salt concentration, and temperature. In paper~\cite{margossian2022coacervation} using similar mean-field theory, the authors described the complex coacervation formed by the liquid-liquid phase separation of a polyzwitterion and a polyelectrolyte. The authors demonstrated the potential of polyzwitterionic coacervates as a tool for pH-triggered release of pharmaceutically active compounds inside the gastrointestinal tract. {In this context, it is valuable to mention several theoretical studies that investigate the complexation of DNA with lipid membranes~\cite{may2000phase,cherstvy2007electrostatics,tarahovsky2009cell,caracciolo2012cationic,cherstvy2014modeling}. Similar to DNA complexation through counterion condensation, which results in attractive effective Keesom interactions, the mobile charges on lipid membranes adjust in response to the negative electrostatic potential created by DNA, thereby triggering the formation of DNA-cationic-membrane complexes. Nowadays, such complexes are commonly used as successful genetic gene-delivery vectors with adjustable pH-dependent release characteristics. Considering that pH levels in cancer cells differ from those in healthy cells, pH-sensitivity becomes essential for precise targeting of malignant cells using these and similar pH-sensitive drug-delivery methods.} It is worth noting that programmable coacervates based on zwitterionic polymers have recently been proposed as promising materials for purifying soft nanoparticles such as liposomes and extracellular vesicles~\cite{capasso2022programmable,paganini2022high}.

In this paper, we will present a molecular statistical theory of self-coacervation in a solution of dipolar polymers. We will properly account for concentration fluctuations, {which are related to the conformational entropy of the chains} within the Edwards-Muthukumar theory~\cite{muthukumar1982extrapolation}, excluded volume interactions within the lattice-based mean-field theory and electrostatic correlations within the random phase approximation (RPA)~\cite{borue1988statistical,budkov2020statistical}. We will describe the phase behavior of a polymer solution for a wide range of electrostatic interaction strength values. {Moreover, we will investigate the behavior of the surface tension of the coacervate-supernatant phase interface, utilizing a simple version of the polymeric density functional theory, as it is an essential factor for successful practical realization of coacervates applications as drug carriers or coatings and packaging materials~\cite{spruijt2010interfacial}.}

\begin{figure}
\includegraphics[width=15 cm]{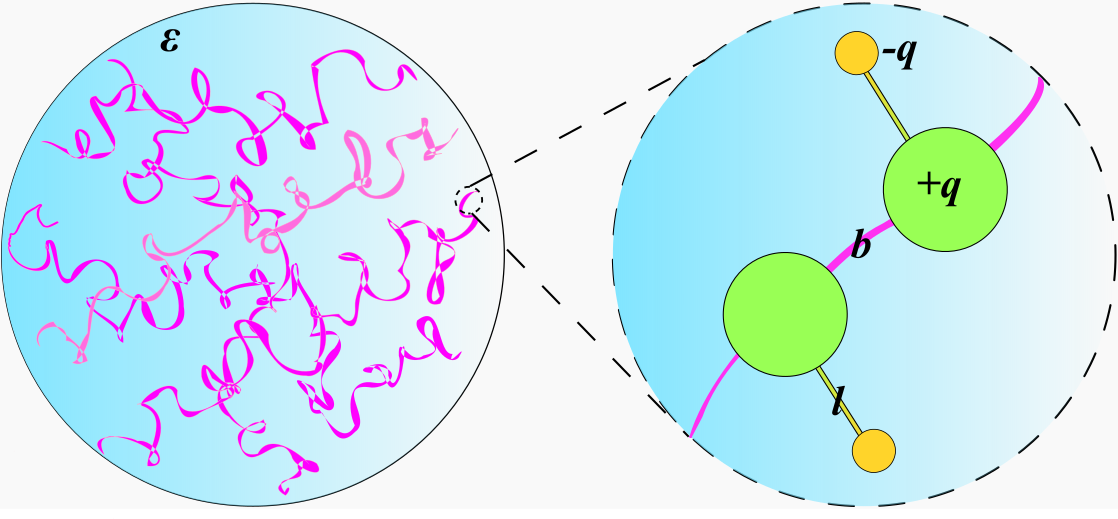}
\caption{Schematic representation of the dipolar polymer solution model.}
 \label{fig0}
\end{figure}

\section{Free energy of a zwitterionic polymer solution}
Let us consider a solution of flexible polymer chains, whose monomeric units carry two oppositely charged sites with charges $\pm q$ separated by a fixed or fluctuating distance described by the probability distribution function, $\omega(\bold{r})$ (the so-called intramolecular correlation function). We also assume that the partial charge $+q$ is located at the monomeric unit, while the point-like partial peripheral charge $-q$ is freely rotating around charge $+q$ (see Fig. 1). We assume that the polymer chains are immersed in a polar solvent with the permittivity $\varepsilon$. The total free energy of such dipolar polymer solution can be written as a sum of two basic terms:
\begin{equation}
f=f_{p}+f_{cor},    
\end{equation}
where
\begin{equation}
\label{f_pol}
\frac{f_{p}}{k_{B}T}=-\frac{9}{32\pi}\frac{vc}{\alpha b^2\xi}+\frac{1}{24\pi \xi^3}+\frac{1}{v}\left((1-cv)\ln(1-cv)+cv\right)+\frac{c}{N}\left(\ln(cv)-1\right) 
\end{equation}
is the free energy density of the basic polymer system taking into account the excluded volume interactions of the monomeric units and their chain connectivity and translation entropy of the polymer chains; $k_{B}$ is the Boltzmann constant, $T$ is the temperature. The first two terms of the interpolation formula (\ref{f_pol}) take into account {the conformational entropy of the flexible chains} within the Edwards-Muthukumar theory~\cite{muthukumar1982extrapolation}. The third term takes into account the excluded volume interactions at a rather high polymer concentration within the lattice model. The forth term is the free energy density of ideal gas of the polymer coils with a polymerization degree $N$. The following values are introduced: chain bond length $b$, concentration of monomeric units $c$, $v$ is the effective volume of a monomeric unit; the coefficient $\alpha$ satisfies the following equation
\begin{equation}
\label{alpha}
\alpha^3-\alpha^2=\pi\sqrt{6}\frac{v\xi}{b^4},
\end{equation}
and the correlation length $\xi$ is determined by the equation
\begin{equation}
\label{xi}
\xi = \frac{1}{2}\left(\frac{9}{16\pi b^2\alpha c}+\sqrt{\left(\frac{9}{16\pi b^2\alpha c}\right)^2+\frac{4b^2\alpha}{cv}}\right).    
\end{equation}
Interpolation formula (\ref{f_pol}) and eqs. (\ref{alpha}-\ref{xi}) allow us to reproduce the well-known scaling expressions for osmotic pressure in the regimes of semi-dilute solution and moderate concentration solution ~\cite{muthukumar1982extrapolation}. In what follows, we assume that $v/b^3=1$, which is a reasonable assumption for the lattice gas reference polymer system. Note that in eqs. (\ref{f_pol}-\ref{xi})
we took into account that the second virial coefficient within the lattice model was $B=v/2$.

The electrostatic correlation contribution to the total free energy density within the RPA has the form~\cite{borue1988statistical,budkov2020statistical}
\begin{equation}
f_{cor}=\frac{k_{B}T}{2}\int\frac{d\bf{k}}{(2\pi)^3}\left(\ln\left(1+\frac{\kappa^2(\bf{k})}{k^2}\right)-\frac{\kappa^2(\bf{k})}{k^2}\right),
\end{equation}
where the screening function
\begin{equation}
\kappa^2({\bf k})=\frac{1}{\varepsilon\varepsilon_0}C(\bold{k})
\end{equation}
with $C(\bold{k})$ being the Fourier-image of the correlation function of the microscopic charge density, $\hat{\rho}(\bold{r})$,
\begin{equation}
C(\bold{r}-\bold{r}^{\prime})=\left<\hat{\rho}(\bold{r})\hat{\rho}(\bold{r}^{\prime})\right>.
\end{equation}
The screening function for zwitterionic polymers, calculated for the first time by two of us (see Supporting information of ref.~\cite{budkov2021molecular}) in the context of the dipolar polymeric gel theory, takes the form
\begin{equation}
\label{screening_function}
\kappa^2({\bf k})=\kappa_{D}^2(1-\omega({\bf k}))\left(1+\frac{1}{2}g(\bold{k})(1-\omega({\bf k}))\right),
\end{equation}
where $\kappa_{D}=r_{D}^{-1}=\left(2q^2c/\varepsilon \varepsilon_0 k_{B}T\right)^{1/2}$ is the inverse Debye length, attributed to the charged groups on the monomeric units and $\omega({\bf k})=\int d{\bf r} e^{-i{\bf k}{\bf r}}\omega({\bf r})$ is the Fourier-image of the probability distribution function, $\varepsilon$ is the solvent dielectric perimittivity, $\varepsilon_0$ is the vacuum dielectric permittivity. The Fourier-image of the pair correlation function of the present basic polymer system in the mean-field approximation is~\cite{de1979scaling,borue1988statistical}
\begin{equation}
\label{pair_cor_func}
g(\bold{k})\simeq\frac{c\xi^3}{1+k^2\xi^2}.
\end{equation}
We would like to note that the semi-dilute regime (\ref{pair_cor_func}) results in the well-known scaling expression for the structure factor~\cite{de1979scaling}, whereas at a rather high polymer concentration it leads to the structure factor expression, obtained in the Edwards concentrated solution regime~\cite{de1979scaling,borue1988statistical}. Then, choosing the function $\omega(\bold{k})=(1+k^2l^2/6)^{-1}$, which quite accurately mimics the case of a stiff dipole of length $l$ allowing us at the same time to obtain the analytical results~\cite{budkov2020statistical,budkov2018nonlocal,budkov2019statistical}, for the case of $\xi \gg l$, where $g(\bold{k})\simeq {c\xi}/{k^2}$, we arrive at
\begin{equation}
\label{f_cor}
f_{cor}=-\frac{k_{B}T}{l^3}\sigma(y)-\frac{k_{B}T}{l^3}\delta(y,\gamma),
\end{equation}
where the first term describes the free energy of disconnected dipolar monomers~\cite{budkov2020statistical,budkov2018nonlocal}, whereas the second one describes the influence of the chain connectivity and the excluded volume interactions on the electrostatic interactions; the auxiliary function~\cite{budkov2019statistical} $\sigma(y)={\sqrt{6}}(4\pi)^{-1}\left(2(1+y)^{3/2}-2-3y\right)$ is expressed in terms of the electrostatic coupling parameter, $y=q^2l^2 c/(3\varepsilon \varepsilon_0 k_{B}T)=4\pi l_{B}l^2c/3$, where $l_{B}=q^2/(4\pi \varepsilon \varepsilon_0 k_{B}T)$ is the standard Bjerrum length. Calculating the correlation length, $\xi$, by eqs. (\ref{alpha}-\ref{xi}) and taking into account that for real macromolecules the relation $l/b \sim 1$ is satisfied, we conclude that $\gamma=c\xi l^2/12$ is rather small in a wide range of polymer concentrations -- from the semi-dilute regime ($cv\ll 1$), where $\xi/b\simeq (9/16\pi)^{3/4}(\pi\sqrt{6})^{-1/4}(cv)^{-3/4}$, to a rather concentrated solution ($cv\sim 1$), where $\xi/b \simeq (3cv)^{-1/2}$. Thus, the cumbersome auxiliary function $\delta(y,\gamma)$ (see Supporting information of ref.~\cite{budkov2021molecular}) can be calculated in the first order approximation in $\gamma$ as follows $\delta(y,\gamma) = 3\sqrt{6}(4\pi)^{-1}\left(1+{y}/{2}-\sqrt{1+y}\right)\gamma + O(\gamma^2)$. Thus, this contribution is negligibly small in a wide range of polymer concentrations, so that it will be safely omitted in practical calculations.

It is instructive to note that the extremely weak influence of the chain connectivity on the electrostatic interactions of the dipolar monomeric units is in accordance with the Lifshitz hypothesis of $"$disconnected$"$ monomers~\cite{lifshitz1978some,khokhlov1994statistical}. More specifically, within this hypothesis, one can consider the contribution of the short-range attractive interactions between the monomeric units (dipole-dipole interaction in our case) to the total free energy to be the same as for the fluid of disconnected monomers.

\section{Coacervation theory}
To analyse the phase behavior of the zwitterionic polymer solution, let us calculate the osmotic pressure
\begin{equation}
\label{Pi}
\Pi =c\frac{\partial{f}}{\partial{c}}-f=\Pi_{p} +\Pi_{cor},
\end{equation}
where the first term in the right hand side of eq. (\ref{Pi}) is the osmotic pressure of the basic polymer system, which has the following asymptotic behavior in semi-dilute and concentrated regimes
\begin{equation}
\label{osm_press}
\Pi_{p}\simeq
\begin{cases}
\frac{ck_{B}T}{N}+\left(\frac{40\sqrt{6}}{81}\right)^{1/4}\pi(cv)^{9/4}\frac{k_BT}{v}, &cv\ll 1\,\\
\frac{ck_{B}T}{N}+\frac{k_{B}T}{v}\left(-\ln(1-cv)-cv\right),&cv \sim 1.
\end{cases}
\end{equation}

The electrostatic correlation contribution to the osmotic pressure not taking into account the influence of the chain connectivity and volume interactions in accordance with the analysis given in the previous section takes the form
\begin{equation}
\label{osm_press_cor}
\Pi_{cor}=-\frac{k_{B}T}{l^3}\sigma_1(y),
\end{equation}
where
$\sigma_1(y)=\sqrt{6}(2\pi)^{-1}\left(3y(1+y)^{1/2}/2-(1+y)^{3/2}+1\right)$ (see ref.~\cite{budkov2020statistical}). At $y\ll 1$ (or $r_D\gg l$), we obtain
\begin{equation}
\label{osm_press_2}
\Pi_{cor} \simeq - \frac{\sqrt{6} q^4lc^2}{48\pi\varepsilon^2\varepsilon_0^2 k_{B}T}=-\frac{k_{B}T}{v}\chi_{K}\phi^2,
\end{equation}
where we have introduced the effective Flory-Huggins (FH) parameter, related to the Keesom interactions 
\begin{equation}
\label{FH}
\chi_{K}=\frac{\sqrt{6} q^4l}{48\pi (\varepsilon \varepsilon_0 k_{B}T)^2v}.   
\end{equation}
In this {\sl Keesom} regime, the electrostatic correlations manifest themselves as effective interactions of point-like freely rotating dipoles -- Keesom interactions~\cite{budkov2020statistical,martin2016statistical,budkov2017polymer} which lead to the second virial coefficient renormalization~\cite{budkov2017polymer,budkov2021molecular}.

In the opposite {\sl Debye} regime, $y\gg 1$ (or $r_D\ll l$), we obtain 
\begin{equation}
\label{Debye}
\Pi_{cor} \simeq  - \frac{k_{B}T}{24\pi r_D^3}.
\end{equation}
In this case, the electrostatic interactions of the charged groups are the same as in polyelectrolyte solutions within the Voorn-Overbeek theory, which in turn is based on the Debye-H{\"u}ckel theory of simple dilute electrolyte solutions~\cite{sing2020recent,overbeek1957phase}. Note that in ref.~\cite{wittmer1993random} similar limiting regimes for the electrostatic correlation contribution to the total free energy (or osmotic pressure) within RPA were obtained for solutions of random and alternating polyampholytes.

\begin{figure}
\includegraphics[width=15 cm]{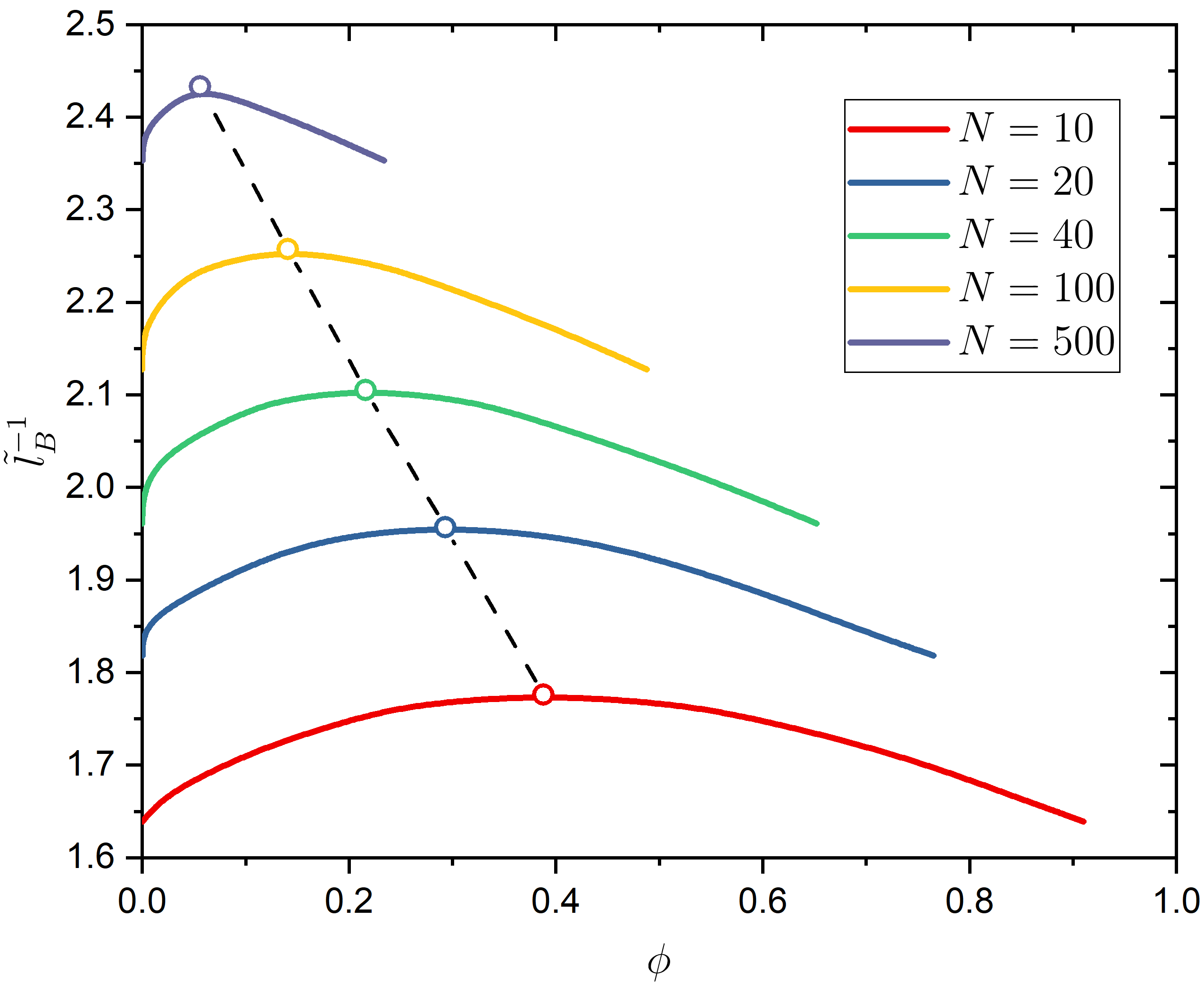}
\caption{{Typical coexistence curves plotted for different polymerization degrees; open circles correspond to the critical points. As is seen, increase in the polymerization degree results in decrease of the polymer concentration in supernatant phase. The data are shown for $\tilde{l}=l/b=0.6$.}}
 \label{binodals}
\end{figure}

\begin{figure}
\includegraphics[width=15 cm]{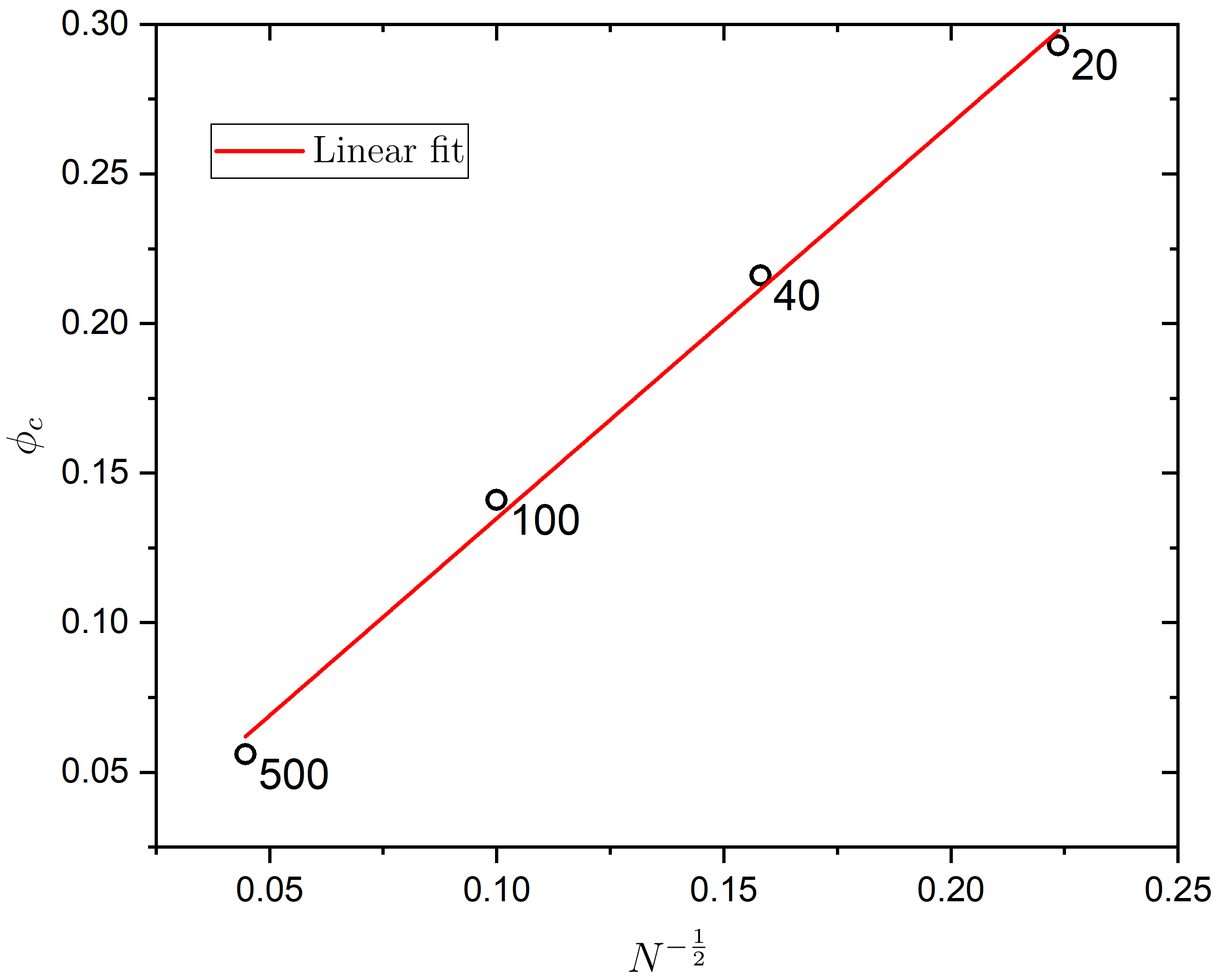}
\caption{{Critical polymer concentrations as a function of inverse square root of the polymerization degree. For convenience the labels near the symbols represent the corresponding degree of polymerisation. The linear fit of the data represents the Flory-Huggins-like trend of the critical concentration behavior.}}
 \label{phi_c_fit}
\end{figure}

As is seen from eq. (\ref{osm_press_cor}), the correlation contribution to the osmotic pressure is always negative, so that at a rather high value of the electrostatic coupling parameter, $y$, the electrostatic correlations can cause the liquid-liquid phase decomposition into two phases -- coacervate phase and supernatant phase. {Fig. \ref{binodals} demonstrates coexistence curves (binodals) for a set of polymerization degrees in $\tilde{l}_{B}^{-1}$-$\phi$ coordinates ($\tilde{l}_B=l_B/b$, $\phi=cv$), computed numerically basing on the standard Maxwell construction~\cite{landau2013statistical}. Note that $\tilde{l}_{B}^{-1}$ can be interpreted as the dimensionless temperature. The curves are plotted for $\tilde l=0.6$ and open circles denote the corresponding critical points. As one can see, with increase in the polymerization degree, the critical concentration decreases, so that at sufficiently large polymerization degree the polymer concentration in supernatant phase becomes negligibly small. The dependence of the critical concentration on the polymerization degree is depicted on Fig. \ref{phi_c_fit}. We obtained that with a good accuracy $\phi_c\approx 1/\sqrt{N}$, that is analogous to the common behavior of the critical concentration, observed in the framework of the standard Flory-Huggins theory~\cite{de1979scaling,khokhlov1994statistical}. In what follows, we will consider only the case of $N\gg 1$ that is realized for polybetaines~\cite{li2020functional}. As is well known, in this case equation $\Pi=0$ allows us to calculate the polymer concentration in the coacervate phase~\cite{borue1990statistical}.}

\begin{figure}
\includegraphics[width=15 cm]{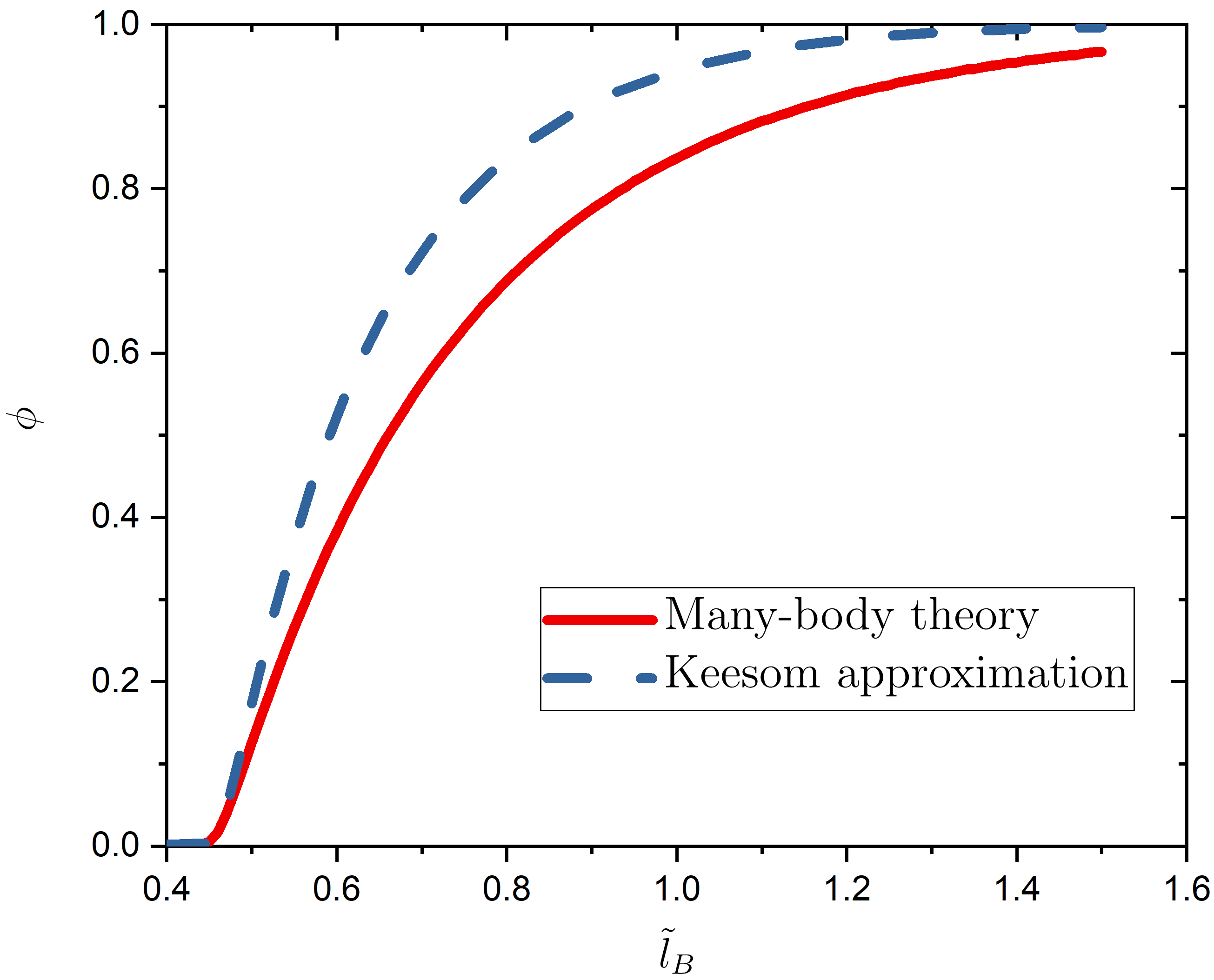}
\caption{Polymer volume fraction, $\phi=cv$, in the coacervate phase as a function of the electrostatic strength, $\tilde{l}_B=l_{B}/b$, calculated within the present theory (solid line) and theory taking into account the electrostatic correlations via effective Keesom interactions (dashed line). The Keesom approximation significantly overestimates the coacervate polymer volume fraction relative to the theory taking into account the electrostatic correlations at the many-body level within RPA. The data are shown for $\tilde{l}=l/b=0.7$.}
 \label{fig2}
\end{figure}

Fig.~\ref{fig2} shows a typical dependence of the polymer volume fraction, $\phi=cv$, in the coacervate phase on the electrostatic interaction strength, $\tilde{l}_B=l_B/b$, for the dipole length $\tilde{l}=l/b=0.7$ and comparison with the result predicted by the theory taking into account the electrostatic correlation via the effective FH parameter~(\ref{FH}) (Keesom approximation). Note that the latter approximation was adopted in recent works~\cite{adhikari2018polyelectrolyte,margossian2022coacervation}. As is seen, the theory based on the Keesom approximation coincides with the present RPA-based theory only within a narrow region of the electrostatic strength (where the polymer volume fraction in the coacervate is close to zero). However, at a rather high value of the electrostatic strength, the Keesom approximation considerably overestimates the polymer volume fraction in the coacervate phase relative to the RPA-based theory. The latter is related to the fact that the Keesom approximation overestimates the contribution of the correlation attraction of zwitterionic monomeric units to the total free energy.

\begin{figure}
\includegraphics[width=15 cm]{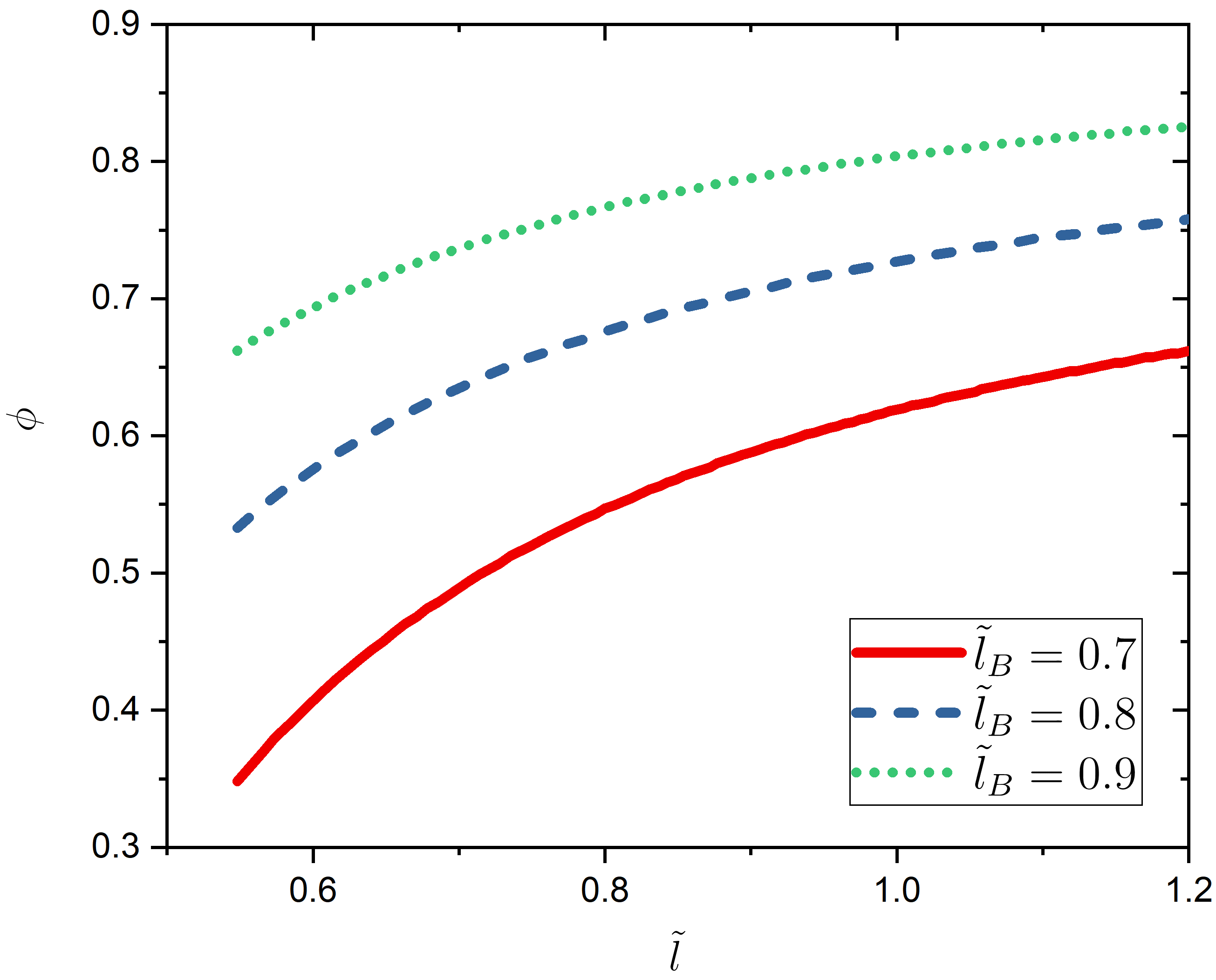}
\caption{Polymer volume fraction, $\phi=cv$, in the coacervate phase as a function of the dimensionless length, $\tilde{l}=l/b$, calculated for different electrostatic strengths.}
 \label{fig3}
\end{figure}

We would like to note that the polymer volume fraction in the coacervate monotonically increases with the growth in the dimensionless dipole length $\tilde{l}$. Such behavior can be interpreted as follows: the stronger the electrostatic interactions of zwitterionic monomeric units, the denser the coacervate phase (see Fig. \ref{fig3}). It is also interesting to note that for denser coacervates (stronger electrostatics), this dependence is weaker (closer to horizontal). Such behavior can be explained by the fact that electrostatic correlations in a denser coacervate are closer to the Debye regime described above, in which the dipole length drops out from the electrostatic free energy expression (see eq. (\ref{Debye})).
Note that such dependences of the coacervate concentration on the dipole length, which are quite transparent from the standpoint of physics, could be used by chemists to synthesize a zwitterionic polymer with a predetermined dipole length to obtain a coacervate with a predetermined density. In this respect, we would like to discuss the parameters of real polymers, for which one can expect to observe coacervation experimentally. For instance, the dipole length of zwitterionic polymers can reach $l\approx 0.5~nm$. Assuming that $v=b^3=1~nm^3$ we obtain a quite reasonable concentration in the coacervate phase $c\approx 0.8~M$ ($\phi\approx 0.5$) for an aqueous solution at $T=300~K$, $\varepsilon=78$, i.e. for $\tilde{l}_{B}\approx 0.7$. Such an estimate should be considered rather rough because the segment length (and effective monomer volume) for real polybetaines with complex molecular monomeric units must be clarified by experimental data treatment.

\section{Surface free energy of a coacervate}
In addition to the bulk coacervate concentration, it is important to calculate the surface tension (or surface free energy) of the coacervate-solvent interface. To achieve this, we can use a simple version of the polymeric density functional theory (DFT) based on the ground state dominance approximation for the conformational free energy~\cite{de1979scaling,khokhlov1994statistical} and local density approximation for the free energy density of the volume interactions of the monomeric units (excluded volume and electrostatic interactions, in this case). The thermodynamic potential per unit interface area of the polymer solution describing the coacervate-solvent interface can be written as follows
\begin{equation}
\label{}
\Omega[c(z)]=\int\limits_{-\infty}^{\infty} dz \left(\frac{k_{B}Tb^{2}}{24c(z)}\left(\frac{dc(z)}{dz}\right)^{2}+f(c(z))-\mu_{b} c(z)\right).
\end{equation}
The first term in the integrand is the standard Lifshitz conformational free energy~\cite{khokhlov1994statistical}. The second term is the free energy density of the spatially inhomogeneous polymer solution (see below); $\mu_b$ is the bulk chemical potential of the monomeric units.

We assume that the transition layer between the coacervate bulk and the pure solvent is located near $z=0$. The coacervate phase is at $z\rightarrow-\infty$, whereas the phase of the pure solvent -- at $z\rightarrow\infty$. The concentration $c(z)$ is determined by the minimization of the functional $\Omega$ that is reduced to solving the Euler-Lagrange equation
\begin{equation}\label{}
\frac{\delta\Omega[\bar{c}(z)]}{\delta c(z)}=0,
\end{equation}
where $\bar{c}(z)$ is the equilibrium concentration profile at the coacervate-solvent interface following from the Euler-Lagrange equation solution. Thus the boundary conditions for the polymer concentration are
\begin{equation}
\label{boundary}
\lim_{z\to\infty}c(z)=0,~\lim_{z\to-\infty}c(z)=c_{b},
\end{equation}
where $c_b$ is the bulk polymer concentration of the coacervate phase satisfying the equation $\Pi_b=0$. The surface tension (or surface free energy) can be calculated as $\sigma = \Omega[\bar{c}(z)]$.

In what follows, we will use the following dimensionless variables $\phi(x)=c(x)b^{3}$, $x={z}/{b}$, $\tilde{f}=fb^{3}/k_{B}T$, $\tilde{\mu}_{b}=\mu_{b}/k_{B}T$, $\tilde{\sigma}=\sigma b^2/k_{B}T$, $\tilde{\mu}_{b}=\tilde{f}(\phi_{b})/\phi_{b}$, $\phi_{b}=c_{b}b^{3}$. Thus the dimensionless surface tension is
\begin{equation}\label{25}
\tilde{\sigma}=\int\limits_{-\infty}^{\infty} dx \left(\frac{1}{24\phi}\left(\frac{d\phi}{dx}\right)^{2}+\tilde{f}-\tilde{\mu}_{b} \phi\right).
\end{equation}
Using the substitution $\psi=\sqrt{\phi}$, eq. (\ref{25}) can be reduced to the form
\begin{equation}\label{sigma_dim}
\tilde{\sigma}=\int\limits_{-\infty}^{\infty} dx \left(\frac{1}{6}\left(\frac{d\psi}{dx}\right)^{2}-U(\psi)\right),
\end{equation}
where the auxiliary function
\begin{equation}\label{}
U(\psi)=\tilde{\mu}_{b} \psi^{2}-\tilde{f}.
\end{equation}
The integrand in eq. (\ref{sigma_dim}) does not explicitly depend on $x$, so that the Euler-Lagrange equation has the first integral
\begin{equation}
\label{const}
\frac{1}{6}\left(\frac{d\psi}{dx}\right)^{2}+U(\psi)=E,
\end{equation}
where $E$ is a constant. Eq. (\ref{const}) is nothing but a mechanical equilibrium condition of coacervate/solvent interface~\cite{budkov2023macroscopic}. As it follows from the expression $\tilde{\mu}_{b}=\tilde{f}(\phi_{b})/\phi_{b}$ (which, in turn, follows from the equation $\tilde{\Pi}_b=0$), in the coacervate bulk $U=0$. The polymer concentration in the coacervate is constant, i.e. $d\psi/dx=0$. Thus, as it follows from the boundary conditions, $E=0$. Therefore, substituting the identity (\ref{const}) into expression (\ref{sigma_dim}), we arrive at
\begin{equation}\label{tension}
\tilde{\sigma}=\int\limits_{0}^{\psi_{b}} d\psi \sqrt{-\frac{2}{3}U(\psi)}.
\end{equation}
where $\psi_{b}=\sqrt{\phi_{b}}$.

In what follows, for simplicity, we will use the following expression for the total free energy density
\begin{equation}
\tilde{f}=(1-\phi)\ln(1-\phi)+\phi-\frac{1}{\tilde{l}^3}\sigma(y).
\end{equation}
Then the dimensionless bulk chemical potential has the form
\begin{equation}
\tilde{\mu}_{b}=-\ln(1-\phi_b)+\frac{\sqrt{6}\tilde{l}_{B}}{\tilde{l}}\left(1-\left(1+\frac{4\pi}{3}\tilde{l}_{B}\tilde{l}^{2}\phi_{b}\right)^{\frac{1}{2}}\right),
\end{equation}
where $\tilde{l}=l/b$ and $\tilde{l}_B=l_B/b$. 

\begin{figure}
\includegraphics[width=15 cm]{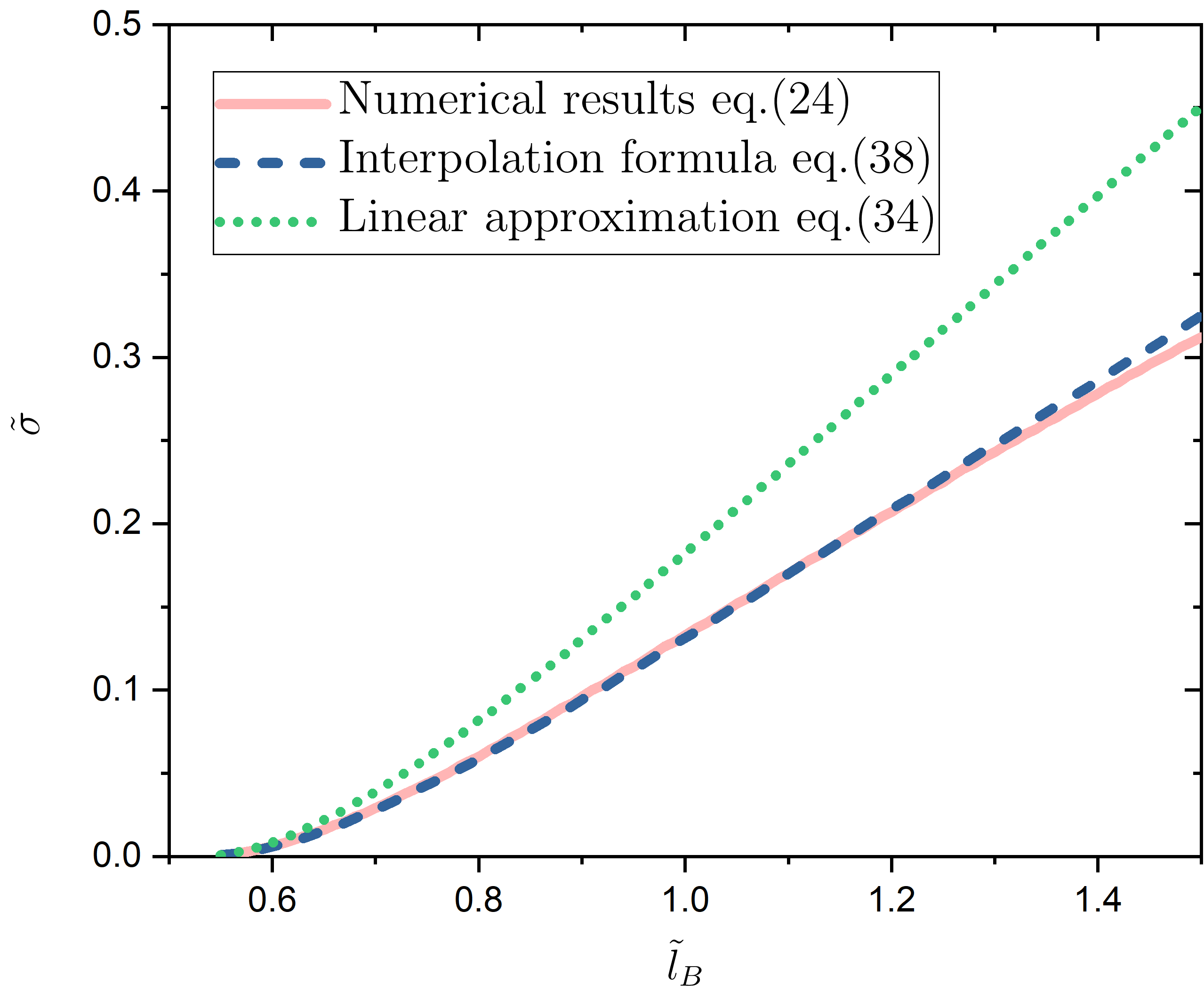}
\caption{Comparison of the surface tension dependences on the electrostatic strength calculated numerically and by approximation formulas. The data are shown for $\tilde{l}=0.7$.}
 \label{fig4}
\end{figure}

In order to analytically estimate the surface tension, we can linearize the equation for the local concentration. For this purpose, firstly, let us divide the transition layer into two regions. In the first one, located near the coacervate bulk, the polymer concentration does not differ much from the bulk one. In the second region corresponding to the transition to pure solvent the local polymer concentration is close to zero. Let us take $x=0$ as the boundary between these two modes. Hereafter one calls these regions the internal and the external boundary layers, respectively. Let us suppose that in the internal boundary layer the function $\psi$ can be represented as
\begin{equation}
\label{psi_pert}
\psi=\psi_{b}+\varphi,\quad \varphi\ll \psi_{b},
\end{equation}
Then the function $U(\psi)$ can be expanded in a series on $\varphi$ truncated by the quadratic term
\begin{equation}
\label{U}
U(\varphi)=-\frac{1}{6}\omega_{0}^{2}\varphi^{2},
\end{equation}
In this case, we obtain
\begin{equation}
\omega_{0}^{2}=\frac{1}{\tilde{D}_{1}^2}=\frac{12\phi_b}{1-\phi_b}
-\frac{8\pi\sqrt{6}\tilde{l}_{B}^{2}\tilde{l}\phi_{b}}{\bigg(1+\frac{4\pi}{3}\tilde{l}_{B}\tilde{l}^{2}\phi_{b}\bigg)^{\frac{1}{2}}},
\end{equation}
where $\tilde{D_1}$ is the dimensionless thickness of the internal layer. An approximate solution to the Euler-Lagrange equation is
\begin{equation}\label{layer1}
\phi=\left(\sqrt{\phi_{b}}-C_{0}e^{\omega_{0}x}\right)^{2},\quad x\leq0,
\end{equation}
Since in the external boundary layer the polymer concentration, $\phi$, tends to zero very fast, we can neglect $\tilde{f}(\phi)$ and keep only the term linear in $\phi$ in the integrand of (\ref{sigma_dim}), which yields
\begin{equation}\label{dif_eq}
\psi^{\prime\prime}=\lambda_{0}^2\psi,
\end{equation}
where the inverse thickness of the external layer
\begin{equation}\label{lambda}
\lambda_{0}=\sqrt{-6\tilde{\mu}_{b}}=\frac{1}{\tilde{D}_{2}}.
\end{equation}
Then, we can derive
\begin{equation}\label{layer2}
\phi=C_{1}^{2}e^{-2\lambda_{0}x},\quad x>0.
\end{equation}
Secondly, from the conditions of equality of functions (\ref{layer1}) and (\ref{layer2}) and their derivatives at $x=0$, we obtain
$C_{1}={\omega_{0}\sqrt{\phi_{b}}}/{(\omega_{0}+\lambda_{0})}$, $C_{0}={\lambda_{0}\sqrt{\phi_{b}}}/{(\omega_{0}+\lambda_{0})}$.
Thus, the surface tension is
\begin{equation}
\label{sigma2}
\tilde{\sigma}=\frac{\phi_{b}}{6\tilde{D}},
\end{equation}
where
\begin{equation}
\tilde{D}=\tilde{D}_{1}+\tilde{D}_{2}
\end{equation}
is the dimensionless thickness of the transition layer.
Expression (\ref{sigma2}) can be rationalized based on the following arguments. Taking into account eqs. (\ref{sigma_dim}) and (\ref{const}), we obtain
\begin{equation}
\tilde{\sigma}=\frac{1}{3}\int\limits_{-\infty}^{\infty} dx\left(\frac{d\psi}{dx}\right)^{2}. 
\end{equation}
Then, taking into account that $|{d\psi}/{dx}|\sim \psi_b/\tilde{D}$ and that the integrand is nonzero only within the transition layer of the thickness $\sim\tilde{D}$, we obtain
\begin{equation}
\tilde{\sigma}\sim \frac{1}{3}\tilde{D}\left(\frac{\psi_b}{\tilde{D}}\right)^2\sim\frac{\phi_{b}}{\tilde{D}}.
\end{equation}

\begin{figure}
\includegraphics[width=15 cm]{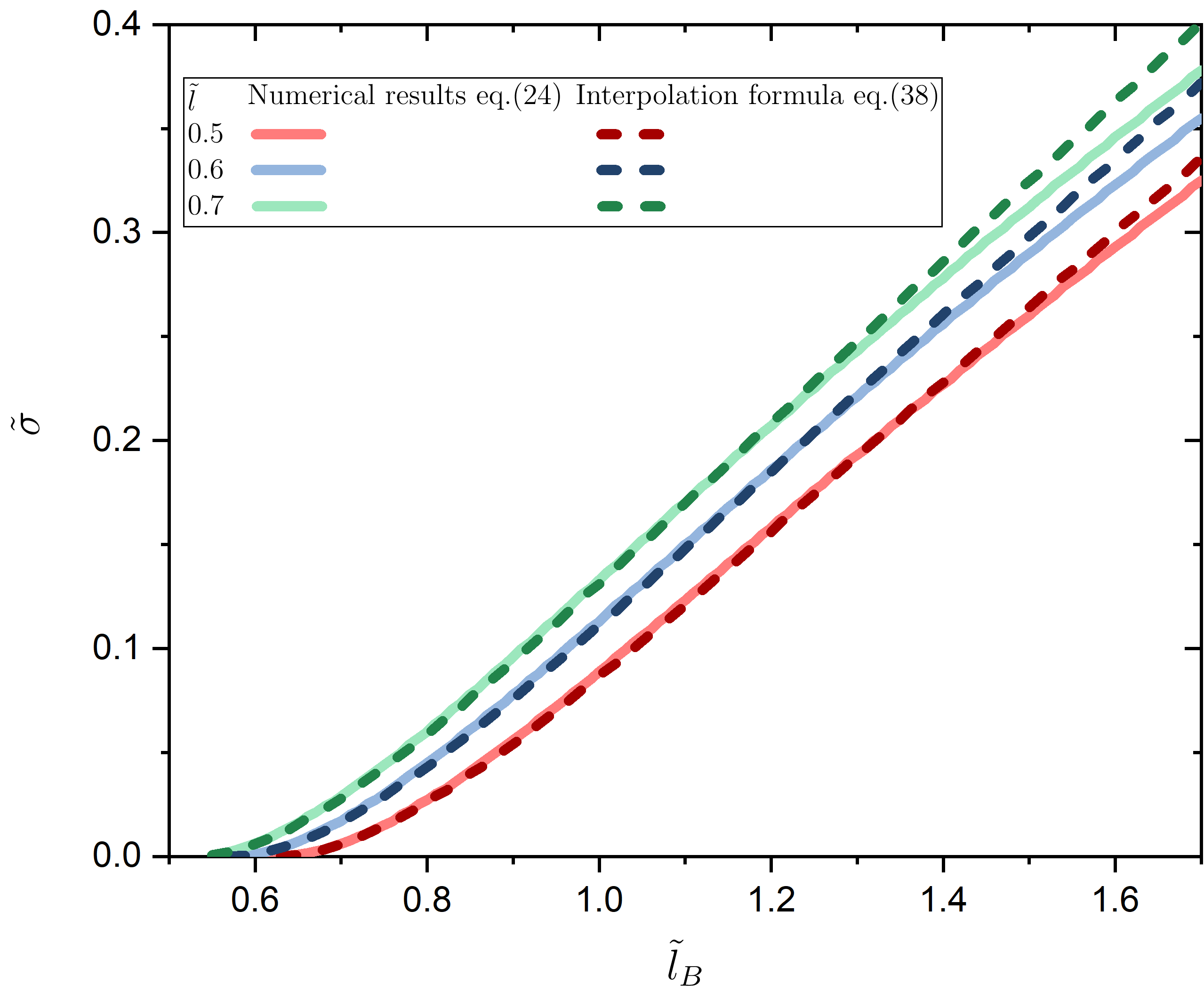}
\caption{Comparison of the surface tension dependences on the electrostatic strength calculated numerically by eq.(\ref{tension}) (solid lines) and approximation formula (\ref{sigma_interpolation}) (dashed lines). With an decrease in the dipole length, the region of electrostatic strength, where the approximation (\ref{sigma_interpolation}) perfectly coincides with the numerical results, expands.}
 \label{fig5}
\end{figure}

As is seen from Fig.~\ref{fig4}, approximation (\ref{sigma2}) gives rather large discrepancy relative to the numerical results obtained by eq. (\ref{tension}). This is related to the fact that dividing the transition layer into two regions seems to be a rather rough approximation to determine the transition layer thickness, $\tilde{D}$. The latter is determined up to a numerical factor of the order of unity. This is justified by the surprise fact that the replacement of factor $1/6$ with $0.12$ in eq. (\ref{sigma2}) gives perfect agreement with the numerical result in a wide range of the electrostatic strength values (see Fig.~\ref{fig4}). A considerable deviation from the numerical results (exceeding 3 \%) in this case is observed only at sufficiently high electrostatic strength values $\tilde{l}_B > 1.2$. However, this region of the electrostatic strength is irrelevant for aqueous zwitterionic coacervates for which $\tilde{l}_B< 1$ or $\tilde{l}_B\sim 1$. Thus, in practical estimations, for aqueous zwitterionc polymer solutions, instead of eq. (\ref{tension}), we can use the following interpolation formula
\begin{equation}
\label{sigma_interpolation}
\tilde{\sigma}=0.12 \times\frac{\phi_b}{\tilde{D}}.
\end{equation}
To calculate the surface free energy of zwitterionic polymer coacervates with organic solvents (alcohols and their aqueous mixtures, for instance), for which the electrostatic strength can reach values higher than unity due to smaller dielectric constants than those of aqueous ones, it is necessary to use eq. (\ref{tension}).

As is seen from Fig.\ref{fig5}, the decrease in the dipole length slightly widens the region of the electrostatic strength where approximation (\ref{sigma_interpolation}) perfectly coincides with the numerical results. We can also see that the surface tension behaves similarly to the polymer volume fraction of the coacervate -- it increases with the dipole length.

\section{Conclusions}
To sum it up, we developed a molecular theory of self-coacervation in semi-dilute zwitterionic polymer solutions. We showed that the competition of the excluded volume interactions of the monomeric units and electrostatic correlations of zwitterionic groups can cause a liquid-liquid phase separation. We calculated the polymer concentration of the coacervate phase depending on the electrostatic interaction strength. We established that in a wide range of polymer concentration values, the chain connectivity and excluded volume interactions of the monomeric units do not affect the contribution of electrostatic interactions of the dipolar monomeric units to the total free energy. In other words, the electrostatic correlation contribution to the total free energy can be described with good accuracy by the expression that is realized for solution of unbounded dipolar monomers. The latter is in accordance with the well-known Lifshitz hypothesis. We showed that for rather low electrostatic strength, the electrostatic correlations manifest themselves as Keesom interactions of point-like freely rotating dipoles (Keesom regime), while in the region of strong electrostatic interactions the electrostatic free energy is described by the Debye-H{\"u}ckel limiting law (Debye regime). We showed that the Keesom regime for real zwitterionic coacervates is realized only for sufficiently low polymer concentrations of the coacervate phase, while the Debye regime is approximately realized for rather dense coacervates. Using the mean-field variant of the density functional theory, we calculated the surface tension (surface free energy) of the $"$coacervate-solvent$"$ interface as a function of the bulk polymer concentration in the definite integral form. We derived the interpolation formula for surface tension which is in good agreement with the numerical results. {Theoretical results that have been obtained in this paper can be utilized to develop more sophisticated models of zwitterionic self-coacervation (for instance, with the add of salt or some target macromolecules to the system). Furthermore, these results can help to estimate the parameters of the polymer chains needed for practical applications such as drug encapsulation and delivery, as well as the design of adhesive materials.}

{\sl Acknowledgements.}  
{\bf YAB dedicates this article to the memory of Igor Ya. Erukhimovich, an outstanding theoretical physicist who had a great influence on him.} The author would like to thank the reviewers for highly significant comments and proposals. YAB and NNK thank the Russian Science Foundation (Grant No. 22-13-00257) for the financial support.

\bibliographystyle{aipnum4-2}
\bibliography{references}

\end{document}